\documentclass[
aps,
pra,
showpacs
amsmath,
amssymb,
preprint,
floatfix,
lengthcheck, 
superscriptaddress,
longbibliography
]{revtex4-1}

\usepackage{graphicx}
\usepackage{mathtools}
\usepackage{braket}
\usepackage{hyperref}
\usepackage{bbold}
\usepackage{calrsfs}
\usepackage{dsfont}
\usepackage{cancel}
\usepackage{mathrsfs}
\usepackage[margin=0.75in]{geometry}
\usepackage[normalem]{ulem}

\usepackage{color}

\DeclareMathAlphabet{\pazocal}{OMS}{zplm}{m}{n}

\makeatletter \renewcommand\@make@capt@title[2]{%
\@ifx@empty\float@link{\@firstofone}{\expandafter\href\expandafter{\float@link}}%
\sffamily{\textbf{#1}}\@caption@fignum@sep#2 }
\usepackage{placeins} 

\definecolor{nrppurple}{RGB}{128,0,128}
\usepackage{hyperref}
\hypersetup{
colorlinks=true,
	linkcolor=nrppurple,
	citecolor=nrppurple,
	urlcolor=nrppurple,
}

\begin{document}


\title{Tunable Nonlinearity and Efficient Harmonic Generation \\ from a Strongly Coupled Light-Matter System}

\author{Davis M. Welakuh}
\email[Electronic address:\;]{dwelakuh@seas.harvard.edu}
\affiliation{Harvard John A. Paulson School Of Engineering And Applied Sciences, Harvard University, Cambridge, Massachusetts 02138, USA}

\author{Prineha Narang}
\email[Electronic address:\;]{prineha@seas.harvard.edu}
\affiliation{Harvard John A. Paulson School Of Engineering And Applied Sciences, Harvard University, Cambridge, Massachusetts 02138, USA}


\begin{abstract}
Strong light-matter coupling within electromagnetic environments provides a promising path to modify and control chemical and physical processes. The origin of the enhancement of nonlinear optical processes such as second-harmonic and third-harmonic generation (SHG and THG) due to strong light-matter coupling is attributed to distinct physical effects which questions the relevance of strong coupling in these processes. In this work, we leverage a first-principles approach to investigate the origins of the experimentally observed enhancement of resonant SHG and THG under strong light-matter coupling. We find that the enhancement of the nonlinear conversion efficiency has its origins in a modification of the associated nonlinear optical susceptibilities as polaritonic resonances emerge in the nonlinear spectrum. Further, we find that the nonlinear conversion efficiency can be tuned by increasing the light-matter coupling strength. Finally, we provide a general framework to compute the harmonic generation spectra from the displacement field as opposed to the standard approach which computes the harmonic spectrum from the matter-only induced polarization. Our results address a key debate in the field, and pave the way for predicting and understanding quantum nonlinear optical phenomena in strongly coupled light-matter systems.
\end{abstract}

\maketitle


\section{Introduction}
\label{sec:introduction}

Nonlinearity, especially higher harmonic generation, is a phenomenon of great fundamental importance. Technologically, harmonic generation is a key mechanism to generate high frequencies for electronics and optoelectronics, enable lasers across broad wavelength ranges, and allow single and pair photon generation for quantum optics. Conventionally, the nonlinear optical process of harmonic generation is a physical effect that occurs when an atomic, molecular, or solid state system is driven by a strong field and emits photons at frequencies which are integer multiples of the fundamental (driving) frequency~\cite{corkum1993,boyd1992}. This multi-photon process has been investigated for several decades and been widely exploited for various technologically-relevant applications~\cite{agrawal2012,zipfel2003,hell2003, klemke2019,gorlach2020}. High-order harmonic generation is usually less efficient compared to low-order (for example, second-harmonic, third-harmonic), due to the weak intrinsic response of the higher-order nonlinear processes. With a wide range of applications of low-order harmonic generation there has been growing interest to diversify ways for efficient  harmonic generation~\cite{saynatjoki2017,shree2021,yao2021,calafell2021,rivera2019a}. 

A promising avenue for efficient harmonic generation is by exploring the phenomenon of strong light-matter interaction within electromagnetic environments such as high-Q optical cavities, nanostructures or plasmonic devices~\cite{flick2018a,ruggenthaler2017b}. The pertinent feature of strong light-matter interaction is the emergence of new hybrid light-matter states known as polaritons which have been shown to influence physical and chemical processes such as modified chemical reactivity~\cite{hutchison2012,schaefer2021}, enhanced conductivity~\cite{orgiu2015}, polariton lasing~\cite{cohen2010}, among other examples. Recently, the field has investigated the influence of polaritons on nonlinear optical processes. For example, strong light-matter interaction within optical cavities have been explored to identify 2p excitons for terahertz lasing devices~\cite{schmutzler2014}, polariton assisted down-conversion~\cite{welakuh2021,juan2020}, frequency conversion for a strongly coupled system~\cite{kockum2017}, efficient second-harmonic generation (SHG)~\cite{chervy2016,pimenta2018}, and third-harmonic generation (THG)~\cite{barachati2018,liu2019} from polaritonic states. Also, using nanoplasmonic devices there has been work aimed at efficient SHG~\cite{kuo2014,ren2014,raygoza-sanchez2019,li2020,li2021}. These examples show the different aspects of nonlinear optics that can be engineered, optimized, or controlled under conditions of strong light-matter interaction.

An intrinsic property of importance to nonlinear optical processes is the nonlinear optical susceptibility which is determined by the microscopic structure of the nonlinear medium. The possibility of changing or controlling this response would lead to significant enhancement of nonlinear optical processes like the case of harmonic generation from polaritonic states~\cite{chervy2016,pimenta2018,barachati2018,liu2019}. However, there is a debate around the origin of the enhancement of SHG and THG from polaritonic states as observed in experiments~\cite{chervy2016,liu2019,barachati2018}. The authors in Refs.~\cite{chervy2016,liu2019} argue that the enhancement has its origins from a genuine
modification of the associated nonlinear susceptibilities while Ref.~\cite{barachati2018} attributes it to the pump field enhancement alone. Since the enhancement of nonlinear effects is of great importance for technological applications across both physical~\cite{boyd1992,agrawal2012} and biological~\cite{zipfel2003,hell2003} sciences, it is important resolve the discussion on the origin of the enhancement of SHG and THG from polaritonic states which show great promise in such applications. For an accurate study, it is important to employ first-principles approaches that can describe the complex interaction when light and matter strongly interact. Such first-principles methods have been developed within the framework of non-relativistic quantum electrodynamics~\cite{ruggenthaler2014,flick2015,flick2018,haugland2020,haugland2021,jestaedt2020,svendsen2021}. Among these methods, quantum electrodynamical density-functional theory (QEDFT) is a valuable approach for describing ground- and excited-states properties of complex matter systems coupled to photons~\cite{flick2017c,flick2019,flick2020,welakuh2022,wang2021,wang2021a,sidler2021,schaefer2021,welakuh2022a} and suitable to investigate nonlinear optical processes of strongly coupled light-matter systems.

In this \emph{Article}, we demonstrate a mechanism to modify and control nonlinear optical processes by strongly coupling a material system to a quantized field of an optical cavity. Specifically, we show for the case of SHG and THG that the associated nonlinear optical susceptibilities are modified due to strong coupling, and this leads to an enhancement of the nonlinear conversion efficiency. This study is particularly important as it addresses the origins of the enhancement of SHG~\cite{chervy2016} and THG~\cite{barachati2018,liu2019} from polaritonic states as observed in experiments. Our \emph{ab initio} results indicate that the enhancement of resonant SHG and THG is attributed to a modification of the involved nonlinear optical susceptibilities as polaritonic resonances emerge in the spectrum. In addition, our results show that the nonlinear conversion efficiency of SHG and THG can be tuned within the cavity by increasing the light-matter coupling strength and we find that these processes are more efficient from the lower polariton state as reported in experiments~\cite{chervy2016,barachati2018}. For the single molecule coupled to a quantized field studied here, we find that THG is seven times more efficient from the lower polariton state when compared to that from a bare matter state. Furthermore, we provide a new approach to compute the harmonic spectrum from the photonic degrees of the coupled light-matter system which equally captures the enhancement of SHG and THG from the polaritonic states. This approach is of fundamental importance because it gives direct access to the harmonic spectrum corresponding to the displacement field (that is, electric field plus induced polarization) as opposed to the standard method which computes the harmonic spectrum from the dipole response (that is, induced polarization only). The results presented here lay the groundwork for understanding the nonlinear optical properties of strongly coupled light-matter systems and present a route towards the investigations of the quantum nature of the radiated harmonic field such as its statistical and non-classical properties. Our \textit{ab initio} approach provides access to observables in quantum nonlinear optics and it is applicable to a wider range of nonlinear optical processes such as two-photon absorption, optical Kerr effect, or four-wave mixing.

\section{Theoretical framework}
\label{sec:general-framework}

To describe the dynamics of matter coupled to photons in the non-relativistic setup we assume that the relevant photon modes have wavelengths that are large compared to the microscopic scale of a matter subsystem, thus enforcing the long-wavelength limit~\cite{tannoudji1989,craig1998}. In this setting we consider the length form of the Pauli-Fierz Hamiltonian in dipole approximation~\cite{flick2019,rokaj2017}:
\begin{align} 
\hat{H} &=\sum\limits_{i=1}^{N}\left(\frac{\hat{\textbf{p}}_{i}^{2}}{2m} + v_{\textrm{ext}}(\hat{\textbf{r}}_i)\right) + \frac{e^2}{4 \pi\epsilon_0}\sum\limits_{i>j}^{N}\frac{1}{\left|\hat{\textbf{r}}_i-\hat{\textbf{r}}_j\right|}\nonumber\\
& \quad +\sum_{\alpha=1}^{M}\frac{1}{2}\left[\hat{p}^2_{\alpha}+\omega^2_{\alpha}\left(\hat{q}_{\alpha} \!-\! \frac{\boldsymbol{\lambda}_{\alpha}}{\omega_{\alpha}} \cdot \hat{\textbf{R}} \right)^2\right] , \label{eq:el-pt-hamiltonian}
\end{align}
where the $N$ electrons are described by the electronic coordinates, $\hat{\textbf{r}}_{i}$, and the momentum operator of the electrons by $\hat{\textbf{p}}_{i}$. The total electronic dipole operator is represented by $\hat{\textbf{R}}=\sum_{i=1}^{N} e \, \hat{\textbf{r}}_{i}$ and $v_{\textrm{ext}}(\hat{\textbf{r}}_i)$ represents the external potential of the nuclei. The quantized electromagnetic field is described by a collection of harmonic oscillators which consists of the displacement coordinate $\hat{q}_{\alpha}$ and canonical momentum operator $\hat{p}_{\alpha}$ with associated mode frequency $\omega_{\alpha}$ for each mode $\alpha$ of an arbitrarily large but finite number of photon modes $M$. The physical coupling between light and matter is $\boldsymbol{\lambda}_{\alpha}=\sqrt{1/\epsilon_{0}V}\textbf{e}_{\alpha}$ where $V$ is the quantization volume of the field and $\textbf{e}_{\alpha}$ is the polarization. Treating the photon degrees of freedom on the same quantized footing as the matter system give access to observables not commonly accessible in a matter-only description (i.e. the semi-classical treatment) such as photon occupation, statistics, entanglement and other non-classical signatures like squeezing~\cite{welakuh2021}. Thus, the light-matter interaction described by Eq.~(\ref{eq:el-pt-hamiltonian}) provides a framework rich in not only the modification of physical and chemical properties of the matter subsystem but also of the photonic subsystem.

\begin{figure}[bth] 
\centerline{\includegraphics[width=0.4\textwidth]{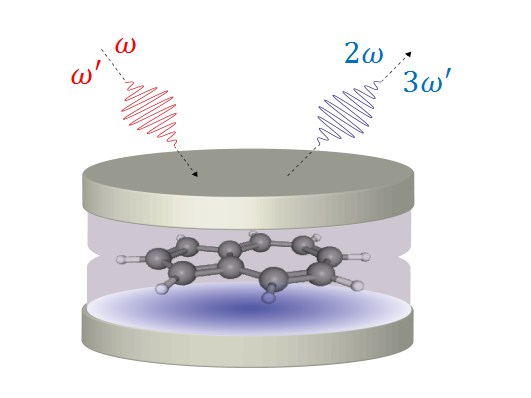}}
\caption{Schematic setup to realize harmonic generation for a strongly coupled light-matter system. An azulene molecule is confined within a high-Q optical cavity and interacts resonantly with a cavity mode. Driving the coupled system with an external field of fundamental frequency $\omega$ ($\omega'$) results to a SHG (THG) at $2\omega$ ($3\omega'$).}
\label{fig:phenol-in-cavity}
\end{figure}

\begin{figure}[bth]
\centerline{\includegraphics[width=0.5\textwidth]{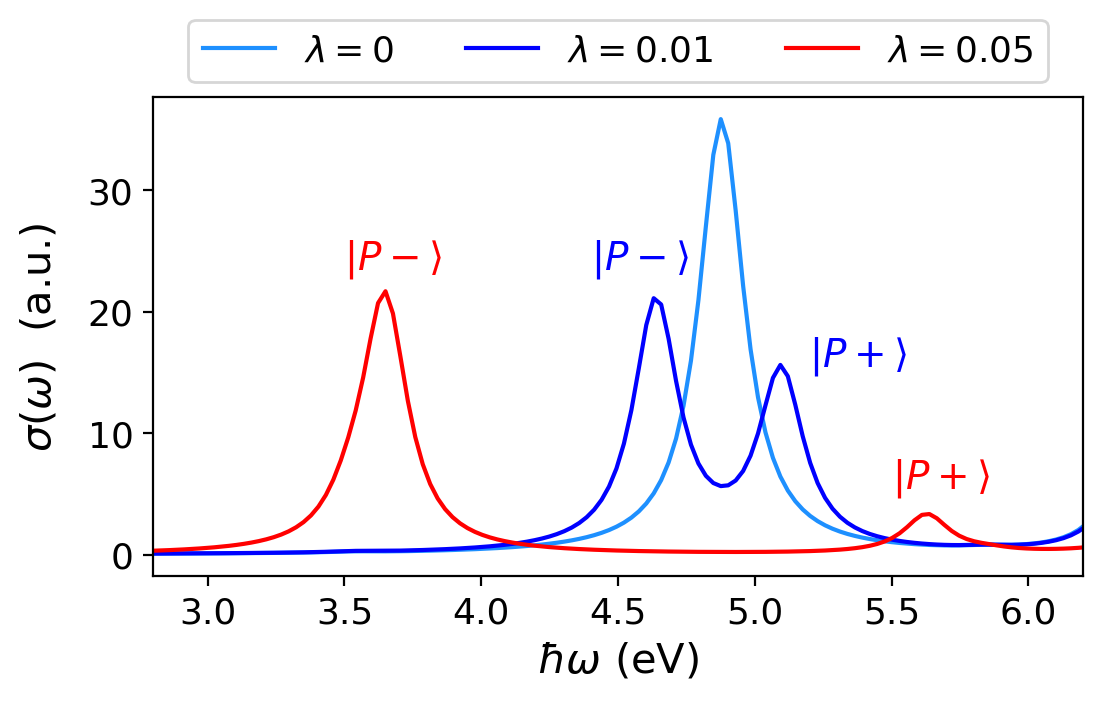}}
\caption{Photo-absorption cross-section of an azulene molecule outside the cavity (when $\lambda=0$) and within an optical cavity (when $\lambda>0$) where a single mode is coupled resonantly to the $\pi-\pi^{*}$ excitation. Varying the coupling strength result to a Rabi splitting into lower and upper polariton peaks with different excitation energies and oscillator strengths.}
\label{fig:azulene-cross-section}
\end{figure}

Next, to investigate how to modify the nonlinear optical susceptibility and hence tune the efficiency of a harmonic generation process due to strong light-matter coupling, we consider the physical setup depicted in Fig.~(\ref{fig:phenol-in-cavity}). A single azulene (C$_{8}$H$_{10}$) molecule is confined within a high-Q optical cavity and interacts strongly with a single photon mode. The azulene molecule has a non-centrosymmetric molecular arrangement which is favorable for a SHG process. In the ultra-violet region of the absorption spectrum of the azulene molecule is a sharp absorption peak at $4.865$~eV that corresponds to the $\pi-\pi^{*}$ excitation energy of the molecule~\cite{welakuh2022}. Since we are interested in harmonic generation from polaritonic states, it is intuitive to compute the spectrum for the photo-absorption cross-section of the coupled system since it captures the hallmark of strong light-matter coupling (i.e. Rabi splitting between polaritons) usually identified in linear spectroscopy. To do this, we confine the molecule within a cavity and set the frequency of the cavity mode in resonance to the $\pi-\pi^{*}$ excitation energy. This leads to a splitting of the peak into lower and upper polariton peaks as shown in  Fig.~(\ref{fig:azulene-cross-section}). From the absorption spectrum of the coupled system, we find new eigenstates (lower polariton $|P-\rangle$ and upper polariton $|P+\rangle$ states) with different excitation energies due to the nonlinear interaction of a cavity photon and the matter system. This already gives us an intuition that when pumping the system, the fundamental frequency of the pump mode for the harmonic generation processes will differ from that outside the cavity since the dynamic dipole polarizability (related to the first-order susceptibility) has been modified due to strong light-matter interaction. The single molecule coupled to a photon mode considered here is relevant to assess the origin of the  enhancement of SHG and THG since the the quantum features of strong light-matter coupling is well understood and easily identified in such a system.

\section{Modified Harmonic generation for a strongly coupled system}
\label{sec:harmonic-generation-polaritons}

We now investigate the possibility of modifying the nonlinear optical susceptibility by strongly coupling a matter system to a cavity mode. Having control over the modification of such intrinsic properties leads to the possibility of enhancing nonlinear optical processes like the SHG and THG. This investigation also resolves the question on the origin of the enhancement of the nonlinear conversion efficiency where Refs.~\cite{chervy2016,liu2019} attribute it to a modification of the associated nonlinear susceptibility and Ref.~\cite{barachati2018} attribute it to a pump field enhancement alone. Our investigation of these nonlinear optical effects employs the accuracy of a first-principles approach based on the real-time propagation framework of QEDFT~\cite{flick2018}. We note that although we consider a planar cavity as shown in Fig.~(\ref{fig:phenol-in-cavity}) to investigate the harmonic generation processes, several other types of electromagnetic environments for achieving strong light-matter interaction like a nanophotonic structure is possible and can be describe with QEDFT~\cite{jestaedt2020}.

\subsection{SHG from polaritonic states}
\label{subsec:SHG-from-polaritons}

\begin{figure*}[bth]
\begin{minipage}[c]{0.51\textwidth}
	\includegraphics[width=3.in,height=2.75in]{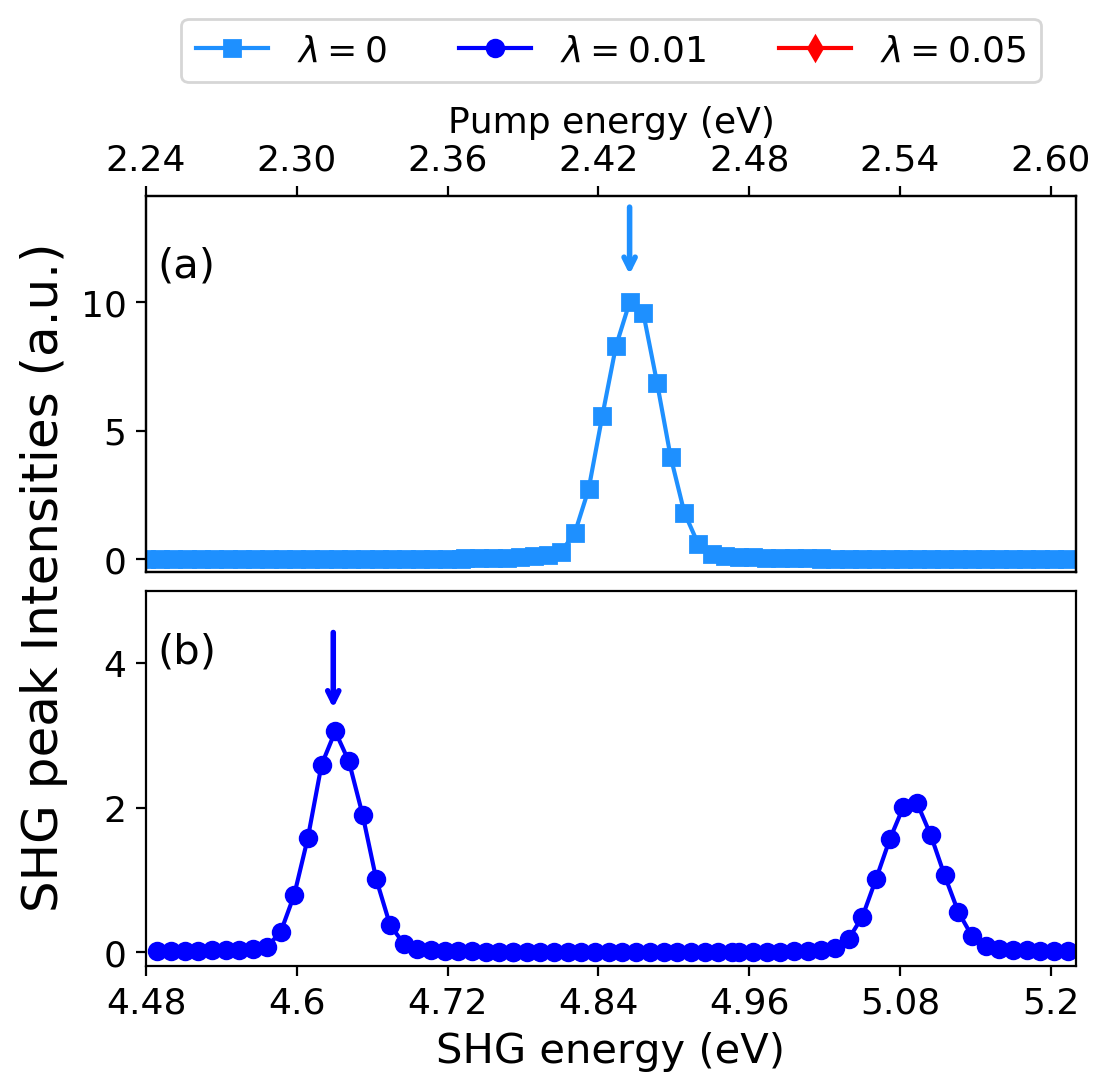}
\end{minipage}%
\begin{minipage}[c]{0.51\textwidth}
	\includegraphics[width=3.in,height=2.15in]{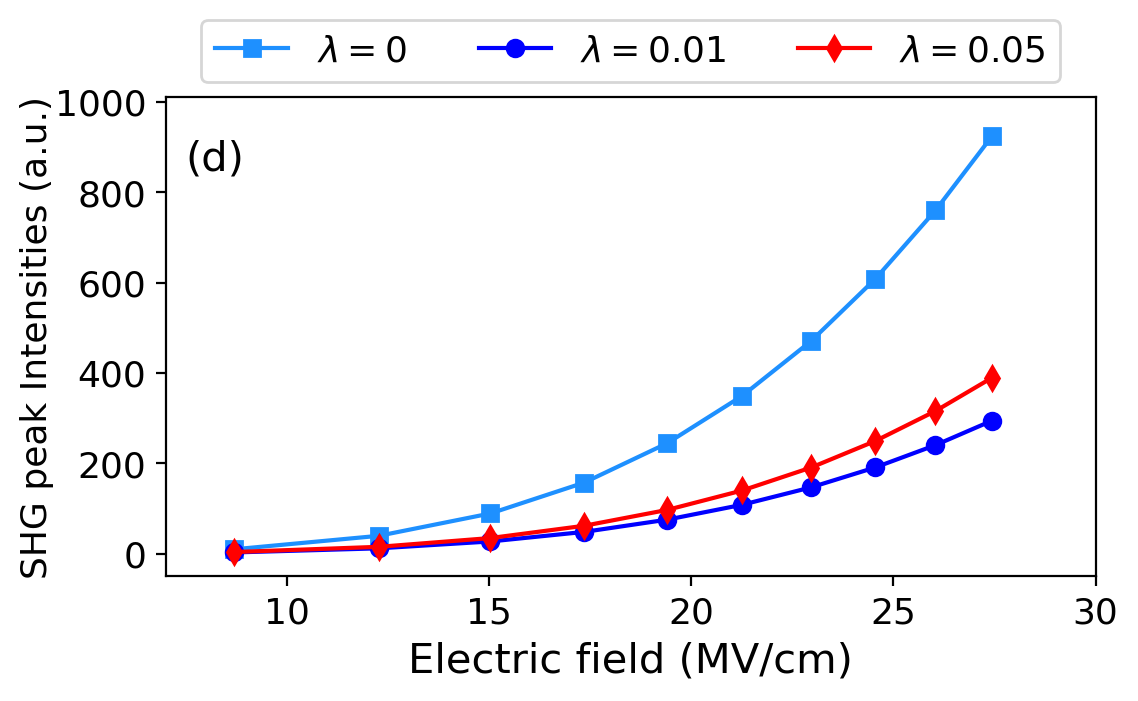}
\end{minipage}\vspace{-15pt}\\
\begin{minipage}[c]{0.51\textwidth}
	\includegraphics[width=3.in,height=2.00in]{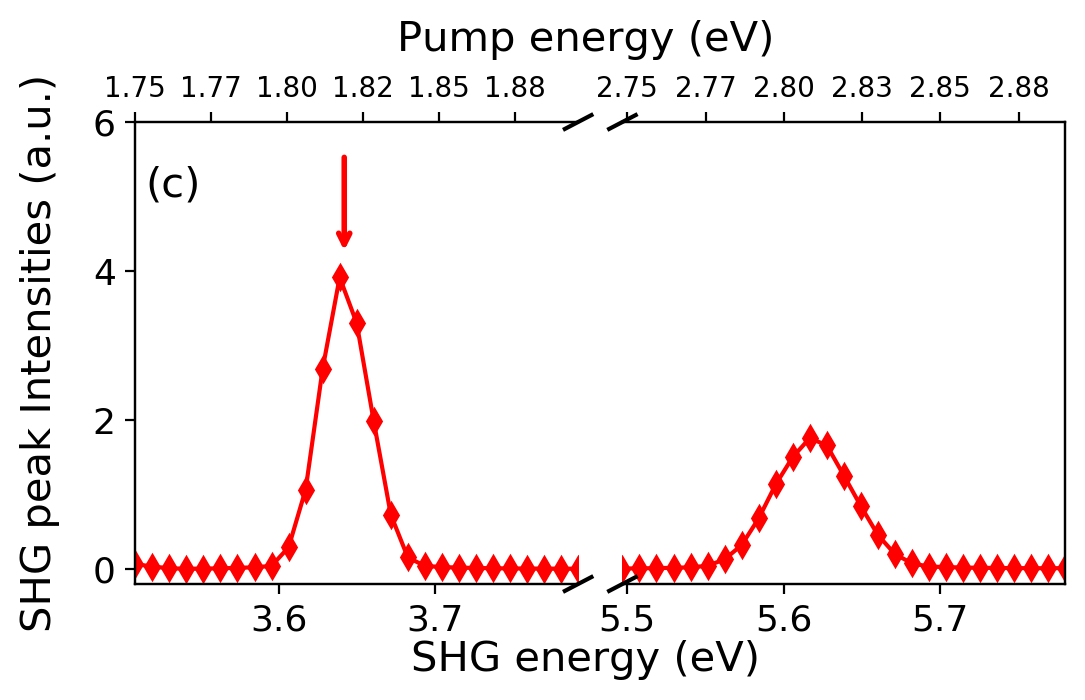}
\end{minipage}%
\begin{minipage}[c]{0.51\textwidth}
	\includegraphics[width=2.98in,height=2.4in]{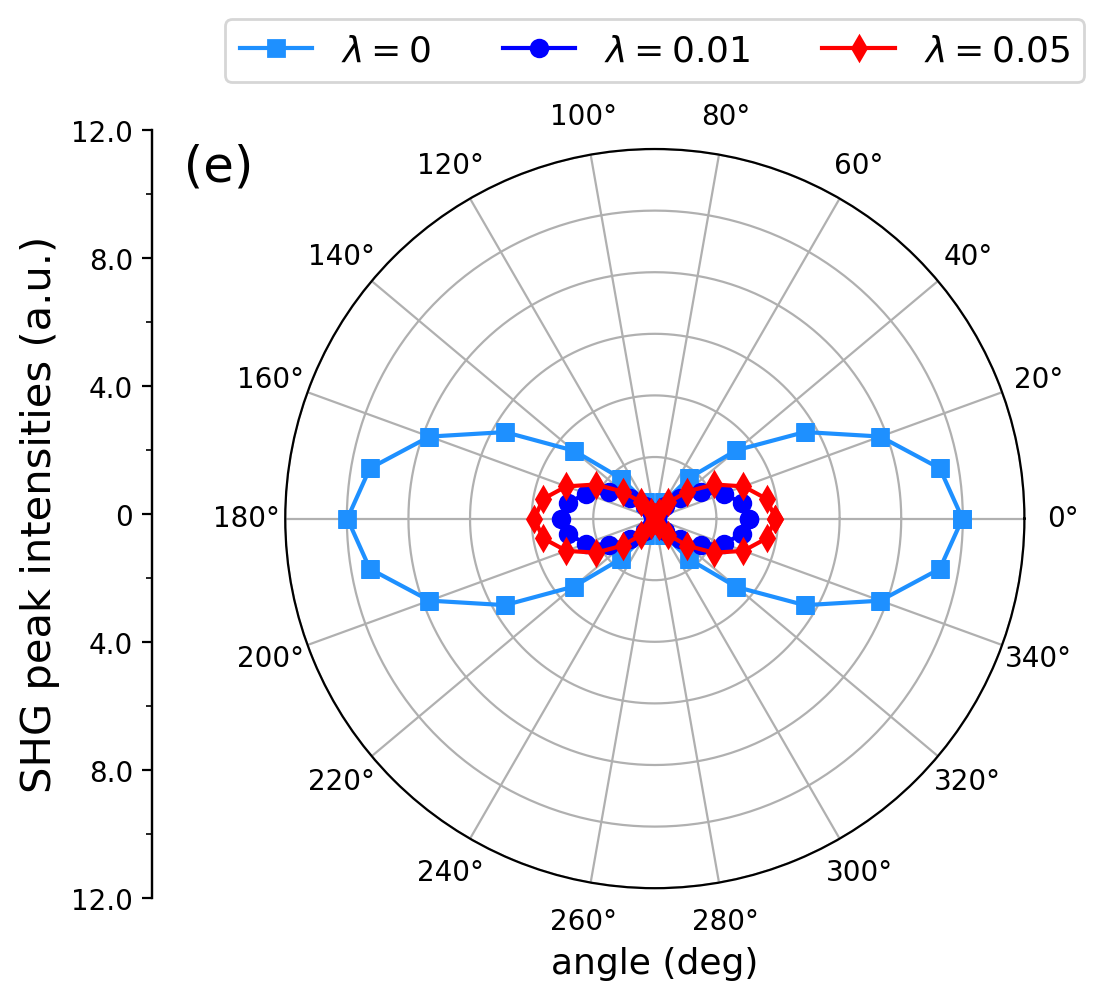}
\end{minipage}
\caption{(a) Calculated second-harmonic spectra for different pump energy for an azulene molecule outside the cavity ($\lambda=0$) which show a central peak (indicated with an arrow) at 4.865~eV. (b) and (c) show the SHG spectra when the molecule is inside the cavity showing two distinct peaks at $4.63$ and $5.09$~eV (for $\lambda=0.01$) and $3.63$ and $5.62$~eV (for $\lambda=0.05$) which originate from the lower and upper polariton states. (d) The SHG intensity dependence as a function of the excitation field, which has a quadratic dependence outside (for $\lambda=0$) and inside (for $\lambda>0$) the cavity for the energy indicated by the arrows. (e) Polar plot of the pump polarization-dependent second-harmonic radiation pattern showing an intensity dependence that fits to $\sim \cos^{2}(\theta)$.}
\label{fig:el-pt-SHG-spectrum}
\end{figure*}

We begin by first investigating the SHG from the azulene molecule outside the cavity (free space where $\lambda=0$). The molecule is driven with a femto-second laser pulse and we vary the fundamental frequency (or pump energy) of the incident field within the range $2.24$ to $2.61$~eV. We compute the second-harmonic field by taking the Fourier transform of the accelerated dipole response (see App.~\ref{app:qedft-harmonic-spectrum} for details). The results of this calculation is shown in Fig.~(\ref{fig:el-pt-SHG-spectrum} a) for $\lambda=0$ where for a pump energy at 2.433~eV, the calculated SHG spectrum has a maximum peak intensity centered at an energy of 4.865~eV. A stringent test to confirm the generated signal is indeed a second-harmonic is to compute its dependence on the electric field amplitude of the pump laser. In Fig.~(\ref{fig:el-pt-SHG-spectrum} d), we show the SHG intensity dependence on the excitation field for the central peak of Fig.~(\ref{fig:el-pt-SHG-spectrum} a) revealing a quadratic dependence (i.e. second-harmonic intensity $\propto \textbf{E}^{2}$) as expected for a SHG process. We also investigate the polarization dependence of the SHG spectra for the central second-harmonic peak of Fig.~(\ref{fig:el-pt-SHG-spectrum} a). The polarization of the incident field is chosen such that the angle $0^{\circ}$ corresponds to the polarization direction along the $x$-axis. In Fig.~(\ref{fig:el-pt-SHG-spectrum} e) we show a polar plot of the calculated SHG intensity dependence on the pump polarization where we find that the process is maximum when the radiated field is parallel to the pump field.

Next, we perform the SHG calculation when the molecule is placed inside an optical cavity and from this we show that the nonlinear optical susceptibility of the molecule gets modified due to strong light-matter interaction. Now, a single mode of the cavity resonantly couples to the $\pi-\pi^{*}$ excitation and we consider two different light-matter couplings $\lambda=0.01$ and $\lambda=0.05$ which can be achieved experimentally by reducing the cavity volume~\cite{barachati2018}. As the linear-response spectrum of the coupled system for these two coupling strengths changes (as in Fig.~(\ref{fig:azulene-cross-section})), we expect that the second-order non-linear response spectrum should also reflect these changes (i.e. polaritonic features also emerge). For the coupled light-matter system, we vary the fundamental frequency of the incident field within the range $1.75$ to $2.89$~eV and compute the energy-dependent SHG spectra for the two cases. The results in Fig.~(\ref{fig:el-pt-SHG-spectrum} b,c) reveal a modified SHG spectrum for the coupled light-matter system. Due to the hybridization and splitting of the $\pi-\pi^{*}$ excitation into lower and upper polariton peaks for different $\lambda$, we no longer find a SHG signal centered at 4.865~eV but find two distinct peaks at $4.63$ and $5.09$~eV (for $\lambda=0.01$) and $3.63$ and $5.62$~eV (for $\lambda=0.05$) which correspond to second-harmonics from the lower and upper polariton states, respectively. The generated signals from the polariton states are of a second-harmonic origin since the harmonic intensities have a quadratic dependence on the field amplitude as shown in Fig.~(\ref{fig:el-pt-SHG-spectrum} d). With the same polarization angle as  the case outside the cavity, we find in Fig.~(\ref{fig:el-pt-SHG-spectrum} e) a similar dipolar SHG emission pattern for the different values of $\lambda$ similar to the case outside the cavity.

\begin{figure}[bth]
\centerline{\includegraphics[width=0.5\textwidth]{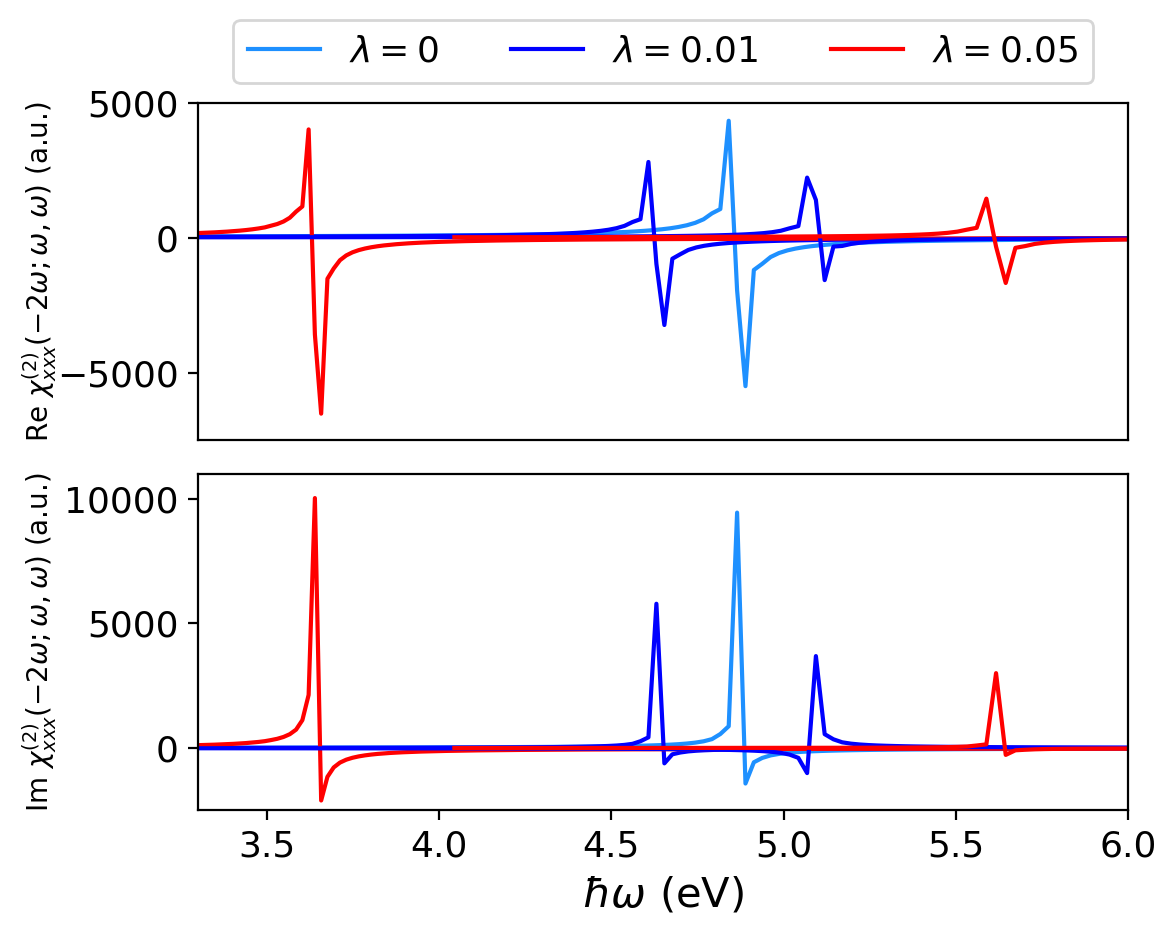}}
\caption{The real and imaginary part of a component of the second-order nonlinear susceptibility tensor of an azulene molecule outside the cavity (when $\lambda=0$) and within an optical cavity (when $\lambda>0$). Varying the coupling $\lambda$ results to a modified nonlinear susceptibility.}
\label{fig:azulene-modified-chi2}
\end{figure}

Notably, the SHG intensity profile from $|P-\rangle$ for different $\lambda$ have higher intensities when compared to that from $|P+\rangle$. This character is reminiscent of the polariton peaks in the absorption spectra (see Fig.~(\ref{fig:azulene-cross-section})) where the peak strengths are directly related to the magnitude of the transition state amplitudes. Just as the dynamical polarizability is modified, the second-order nonlinear susceptibility $\chi_{ijk}^{(2)}(-2\omega;\omega,\omega)$ is also modified when strongly coupled to a photon mode. To make this evident, we compute the second-order nonlinear susceptibility associated to the SHG in Fig.~(\ref{fig:el-pt-SHG-spectrum} a-c). This is done following the methods described in Refs.~\cite{takimoto2007,goncharov2012}. In Fig.~(\ref{fig:azulene-modified-chi2}), we show clearly a modification of a component of $\chi_{ijk}^{(2)}$ for the SHG when the molecule is placed within the cavity as polaritonic peaks emerge. Changing the coupling $\lambda$ allow to control the nonlinear optical properties of the molecule and with it, the nonlinear optical effects. The emergence of polaritonic peaks in $\chi_{ijk}^{(2)}$ accounts for the two distinct peaks in Fig.~(\ref{fig:el-pt-SHG-spectrum} b,c) and highlights that the generated second-harmonic signals has its origins in the strongly coupled light-matter system as observed in experiments~\cite{schmutzler2014,chervy2016,barachati2018,li2020}. Alternatively, the modification of $\chi_{ijk}^{(2)}$ can be inferred from the sum-over-states (SOS) model of nonlinear optics~\cite{kuzyk2013} where for the coupled system, the SOS includes the energies and transition state amplitudes of the polaritons.

\subsection{THG from polaritonic states}
\label{subsec:THG-from-polaritons}

For the THG, the process is weaker than the SHG due to the weak intrinsic response of the third-order nonlinear susceptibility associated with THG. To obtain a larger yield of the THG, the intensity of the  femto-second laser pulse is increased (see App.~\ref{app:azulene-harmonic-generation} for details). The simulations are performed similar to the SHG case except that we vary the fundamental frequency of the pump field to be one-third the excitation energy of the $\pi-\pi^{*}$ transition (for $\lambda=0$) and one-third that of the lower and upper polariton states (for $\lambda>0$). 
\begin{figure}[bth]
\centering
\begin{minipage}[b]{0.45\linewidth}
	\centerline{\includegraphics[width=8.5cm,height=7.5cm]{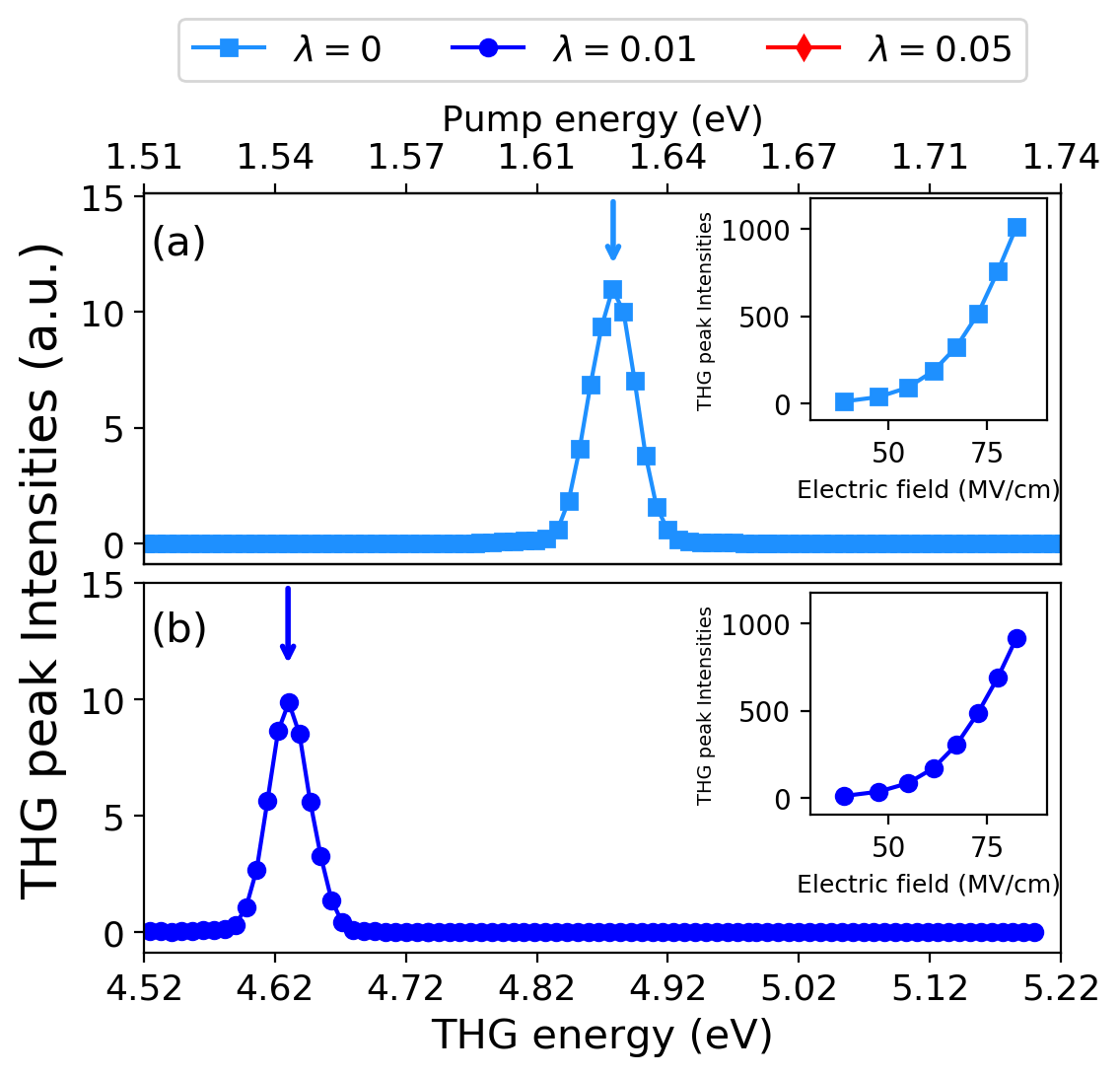}}
\end{minipage}
\\
\begin{minipage}[b]{0.45\linewidth}
	\centerline{\includegraphics[width=8.5cm,height=4.5cm]{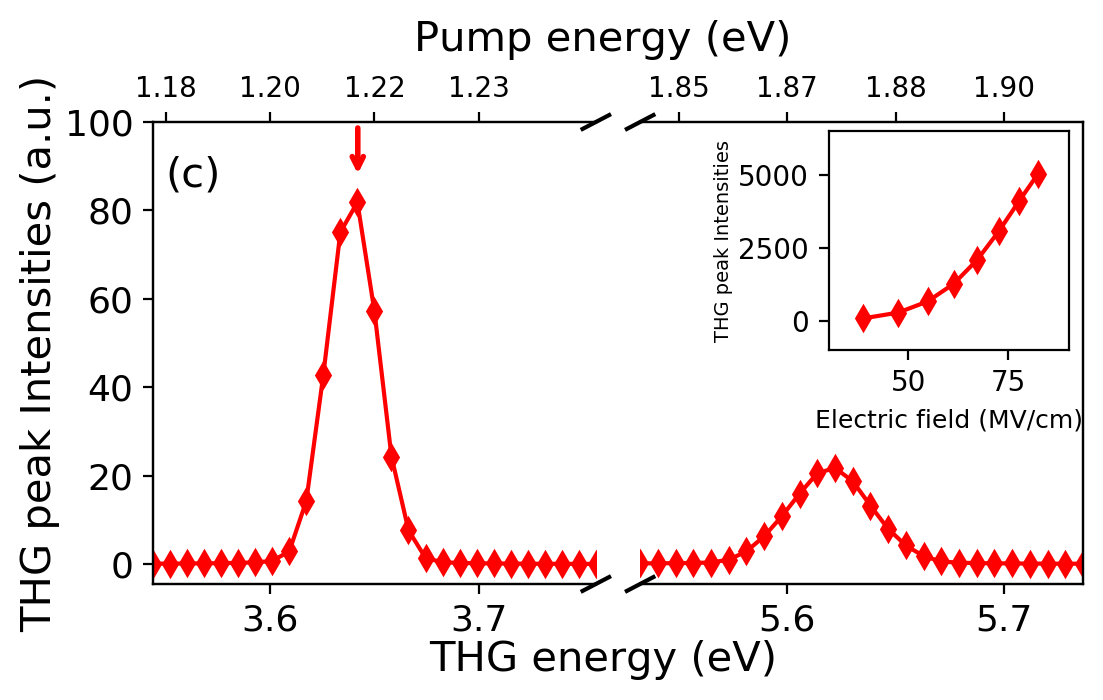}}
\end{minipage}
\caption{THG spectra at different pump energies from an azulene molecule (a) outside ($\lambda=0$) and (b,c) inside ($\lambda>0$) a cavity. (a) THG signals are generated at the energies of the $\pi-\pi^{*}$ transition and polaritonic states but having different intensities. The intensity of the upper polariton in (b) is small with central intensity of $0.014$ (a.u.). The insets in (a-c) show the cubic dependence of the pump field amplitude on the THG intensity computed for the pump energy indicated by the arrows.}
\label{fig:el-pt-THG-spectrum}
\end{figure}
The results of the THG outside and inside the cavity is shown in Fig.~(\ref{fig:el-pt-THG-spectrum}). Outside the cavity, a third-harmonic signal with peak centered at an energy of 4.865~eV is generated for a corresponding pump energy that is one-third the energy of the THG. When placed inside the cavity, a similar spectrum as the SHG case is obtained for different values of $\lambda$ with the only difference being the peak intensities. The lower polaritonic peak is shown to have larger peak intensities when the coupling strength is increased. This finding elucidates the argument that strong coupling leads to an increase yield of the nonlinear process when compared to only strong field pumping~\cite{welakuh2021,welakuh2022c}. The insets in Fig.~(\ref{fig:el-pt-THG-spectrum} a-c) show that the THG intensities have a cubic dependence on the incident field (i.e. THG intensity $\propto\textbf{E}^{3}$) which indicates that we have a THG when the molecule is placed outside and inside the optical cavity. We note that since the resonant THG in Fig.~(\ref{fig:el-pt-THG-spectrum} b,c) correspond to polaritonic peak positions in Fig.~(\ref{fig:el-pt-SHG-spectrum} b,c), this indicates that there is a tunable modification of $\chi_{ijkl}^{(3)}(-3\omega;\omega,\omega,\omega)$ for the strongly coupled light-matter system. This is confirmed in Ref.~\cite{liu2019} where the SOS model was used to obtain a modified $\chi^{(3)}$ in their experiment.

An important aspect of any harmonic generation process is the nonlinear conversion efficiency. To compute the efficiency, we apply the normalized harmonic generation conversion efficiency defined as $\eta = I_{n\omega}/I_{\omega}^{n}$ where $I_{\omega}$ ($I_{n\omega}$) represents the peak intensity of the fundamental frequency (n-th harmonic generation, where $n=2,3,...$)~\cite{ren2014,li2021}. Since we are interested in how the efficiency of the resonant SHG and THG processes inside the cavity compare to that outside the cavity, we define the ratio between their conversion efficiencies: $R=\eta^{(\text{cav})}/\eta^{(\text{no-cav})}$. For a fixed $I_{\omega}$ for SHG (THG) inside and outside the cavity, this ratio becomes: $R=I_{n\omega}^{(\text{cav})}/I_{n\omega}^{(\text{no-cav})}$. We compute this quantity for the central peak intensities of the lower polaritons in Fig.~(\ref{fig:el-pt-SHG-spectrum} b,c) and Fig.~(\ref{fig:el-pt-THG-spectrum} b,c) where we have $R_{\textrm{SHG}}=0.30, \, (0.39)$ and $R_{\textrm{THG}}=0.897, \, (7.449)$ for $\lambda=0.01, \, (0.05)$, respectively. We find that the resonant SHG is not as efficient inside the cavity when compared to that outside the cavity, however, its efficiency is tunable as $R_{\textrm{SHG}}$ increases with increasing $\lambda$. On the other hand, THG is also tunable and it is seven times more efficient inside the cavity than it is outside the cavity for $\lambda=0.05$. We note that $R_{\textrm{THG}}$ is independent of $I_{\omega}$, thus, we cannot attribute the increased efficiency to the increased  pump intensity. As a result, we attribute this increase to a modification of $\chi_{ijkl}^{(3)}$ due to strong light-matter coupling as reported in Refs.~\cite{li2021,chervy2016}. Clearly, changing the coupling strength result to a tunable conversion efficiency which is important for various technological applications across both physical~\cite{boyd1992,agrawal2012} and biological~\cite{zipfel2003,hell2003} sciences.

The \emph{ab initio} investigation of resonant SHG and THG considered here was the single-molecule strong coupling case. In the case that we increase the light-matter coupling into the ultra-strong or deep-strong coupling regime, this will further result to an increase efficiency of the SHG and THG from polaritonic states. For the case of collective strong coupling of an ensemble of molecules where the Rabi splitting is proportional to the square-root of the number of emitters~\cite{sidler2021,sidler2021a}, we expect to get an increased enhancement of these nonlinear processes as reported in experiments for SHG~\cite{chervy2016} and THG~\cite{barachati2018,liu2019}. Nevertheless, the single-molecule strong coupling studied here already sheds light on how the nonlinear conversion efficiency increases for increasing coupling strength. This makes the results presented here the groundwork for understanding nonlinear optical processes of strongly coupled light-matter systems.

\section{Quantum photonic observables in harmonic generation}
\label{sec:photon-SHG}

So far, we employed the standard theoretical approach for computing harmonic generation by modeling the emitted field as a dipole radiation where the dipole field is calculated as an expectation value of the dipole moment. We are however not limited to only electronic observables since the photon field of the cavity is treated on an equal quantized footing as the matter system. Such a treatment provides a plethora of photonic observables that can shed light on the state of the generated harmonic field. For example, we can investigate the statistical and non-classical properties of the nonlinear processes~\cite{gorlach2020,welakuh2021,welakuh2022c}. For the purpose of demonstrating new observables that can be computed from a strongly coupled light-matter system for the harmonic generation process, we present a description of the SHG and THG above by computing the photon displacement coordinate.

\begin{figure}[bth]
\centerline{\includegraphics[width=0.5\textwidth]{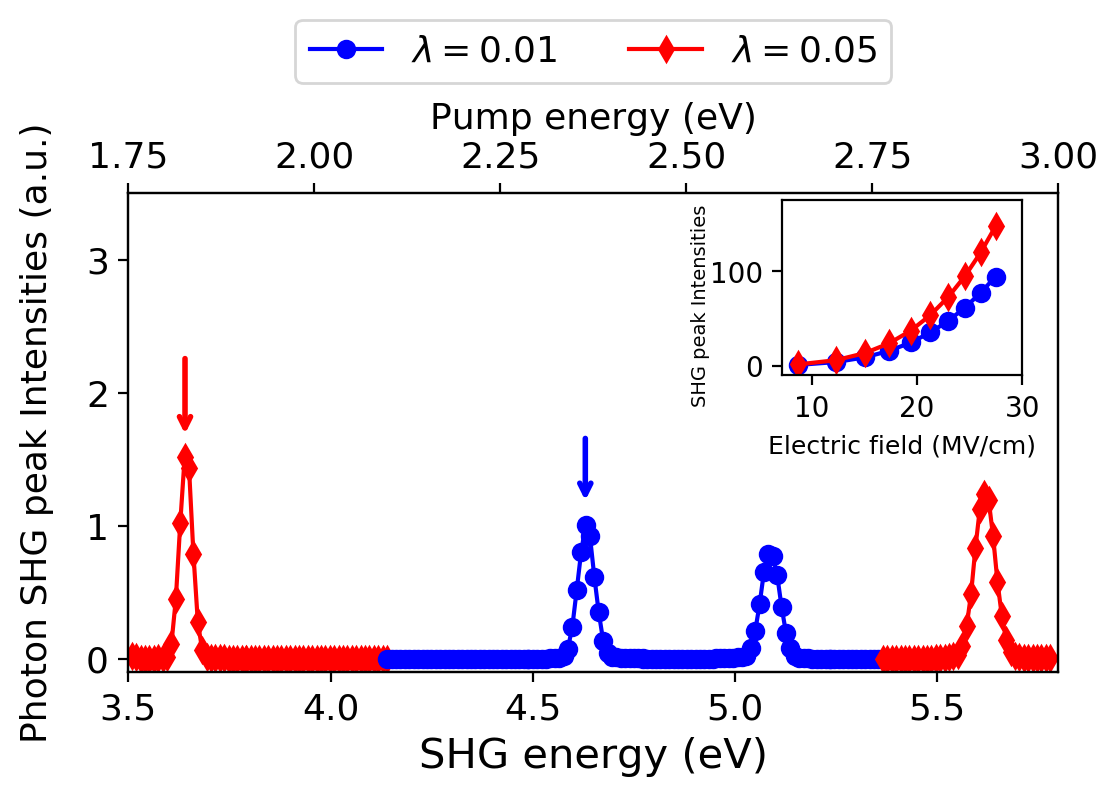}}
\caption{The photon SHG spectrum computed from the displacement field of the coupled system inside the cavity. For different $\lambda$, SHG peaks emerge at the lower and upper polariton states for the respective couplings. The inset show a quadratic dependence of the \textit{photon} SHG intensity for different field amplitudes for the peak intensity indicated with arrows.}
\label{fig:photon-SHG-omega}
\end{figure}

In the length gauge description of the coupled system, the induced electric field is defined as $\hat{\textbf{E}}_{\perp} = \sum_{\alpha} \boldsymbol{\lambda}_{\alpha}\omega_{\alpha}   \left( \hat{q}_{\alpha} - \boldsymbol{\lambda}_{\alpha} \cdot\hat{\textbf{R}}/\omega_{\alpha} \right)$~\cite{rokaj2017}. Since the induced field can be computed directly from this expression, we could as well compute the harmonic spectrum by taking a Fourier transform of the time-dependent field. Here, we will rather compute the harmonic spectrum from the photon displacement coordinate $\hat{q}_{\alpha}$ which is related to the displacement field as 
$\hat{\textbf{D}}=\epsilon_{0}\sum_{\alpha}\omega_{\alpha}\boldsymbol{\lambda}_{\alpha}\hat{q}_{\alpha}$~\cite{rokaj2017}. The reason for this choice is because the observable is a mixed matter-photon quantity in the length gauge and it accounts for the cross-talk between light and matter~\cite{flick2019,welakuh2022}. Also, this observable directly gives the displacement field (electric field plus induced polarization) as opposed to the standard approach which considers only the induced polarization. In analogy to modeling the spectrum from the dipole radiation, we define a similar expression for the power spectrum from the photon displacement coordinate (see App.~\ref{app:qedft-harmonic-spectrum} for details). From this relation we are able to obtain for the first time, a photon perspective of harmonic generation. In Fig.~(\ref{fig:photon-SHG-omega}), we show the \textit{photon} SHG spectra calculated from the photon displacement coordinate. It is important to note that the linear-response spectrum computed from the displacement coordinate equally captures polaritonic features (lower and upper polariton peaks) as shown for the case of the azulene molecule~\cite{welakuh2022}. Thus, it is not surprising to find \textit{photon} SHG originating from polaritonic states as shown in Fig.~(\ref{fig:photon-SHG-omega}). Similar to the SHG in Fig.~(\ref{fig:el-pt-SHG-spectrum} b,c), we have SHG from polaritonic peaks for different light-matter coupling strengths with the only difference being in the peak intensities. The inset in Fig.~(\ref{fig:photon-SHG-omega}) confirms that we indeed have a \textit{photon} SHG from polaritonic states as the second-harmonic intensity has a quadratic dependence on the pump field amplitude. We note that the case $\lambda=0$ is not shown since light and matter decouple. This observable can capture the generation of higher harmonics as we also compute the \textit{photon} THG from polaritonic states as discussed in App.~\ref{app:photon-harmonic-generation}. This new approach is relevant as it captures the important features of enhanced and tunable harmonic generation conversion efficiencies from polaritonic states.  Clearly, treating light-matter interaction on an equal quantized footing opens new avenues for different investigations into photon properties of the generated harmonics such as the photon statistics of the emitted harmonic radiation within the optical cavity~\cite{welakuh2022c}.

\section{Conclusions and Outlook}
\label{sec:conclusion-outlook}

In this work, we show that nonlinearities of matter can be modified and controlled by strongly coupling to a quantized field within an electromagnetic environment. Specifically, we show that the nonlinear optical susceptibility associated with SHG and THG is modified when a photon mode resonantly couples to a matter excitation. By performing an energy-dependent SHG and THG calculation for the molecule inside the cavity, we obtain two distinct SHG and THG signals corresponding to polaritonic states that emerge in the strongly coupled light-matter system. For the coupled system, the nonlinear conversion efficiency can be tuned by increasing the light-matter coupling strength where the conversion efficiency is seven times larger inside the cavity than it is outside the cavity for THG. Since the ratio $R_{\textrm{THG}}$ is independent of the pump intensity, we attribute the increased efficiency to a modification of the nonlinear optical susceptibility as also reported in experiments~\cite{liu2019,chervy2016}. In our system, we find that the resonant SHG and THG is more efficient from the lower polariton state when compared to the upper polariton as observed in experiments~\cite{chervy2016,barachati2018}, warranting further experimental and theoretical investigation. 

Further, we present a general approach to compute the harmonic generation spectra from the photonic degrees which obtains the important physical effects as the standard approach. This new approach differs from conventional methods as it directly computes the harmonic spectrum corresponding to the displacement field as opposed to the standard approach that considers only the induced polarization. The introduced approach is important as it provides information about the photonic properties of the generated harmonics, which are not usually considered. Our theoretical approach and scheme for tuning the efficiency of harmonic generation processes of strongly coupled light-matter systems can be applied to a variety of other systems (atomic as well as condensed matter systems). The \emph{ab initio} method is applicable to a wide range of nonlinear optical processes, paving the way towards novel quantum optical phenomena in nonlinear optics.

Finally, we envision future work focused on investigating the statistical and non-classical properties of the emitted radiation of harmonic generation processes. This is important for applications that rely on the generation and manipulation of non-classical light fields. Another direction of interest would be investigating the possibility to control or modify nonlinear-optical effects such as the optical Kerr effect and THz generation with strong light-matter interactions.

\section{Acknowledgments}

The authors thank Nicolas Tancogne-Dejean, Johannes Flick, Jonathan Curtis, Stephane Kena-Cohen, and Aaron Lindenberg for fruitful discussions, and Andrew Welakuh for graphical support. This work is primarily supported through the Department of Energy BES QIS program on `Van der Waals Reprogrammable Quantum Simulator' under award number DE-SC0022277 for the work on nonlinearity in low dimensional systems, as well as partially supported by the Quantum Science Center (QSC), a National Quantum Information Science Research Center of the U.S. Department of Energy (DOE) on cavity-control of quantum matter. P.N. acknowledges support as a Moore Inventor Fellow through Grant No. GBMF8048 and gratefully acknowledges support from the Gordon and Betty Moore Foundation as well as support from a NSF CAREER Award under Grant No. NSF-ECCS-1944085.

\appendix

\section{Harmonic spectrum from QEDFT}
\label{app:qedft-harmonic-spectrum}

The harmonic spectrum can be determined by solving the time-dependent Schr\"{o}dinger equation with the Hamiltonian of Eq.~(\ref{eq:el-pt-hamiltonian}). Since solving the Schr\"{o}dinger equation incurs huge computational cost even for the simplest systems, we employ the Maxwell-Kohn-Sham construction of QEDFT~\cite{ruggenthaler2014,flick2015,flick2018} that greatly reduces the computational cost and makes it possible to deal with multi-electron systems coupled to photons. The coupled non-linear Maxwell-Kohn-Sham equations for auxiliary electronic orbitals, which sum to the total density $n(\textbf{r},t) = \sum_{i} |\varphi_i(\textbf{r},t)|^2 $, and the displacement fields $q_{\alpha}(t)$ is given by
\begin{align}
i\hbar \frac{\partial}{\partial t} \varphi_{i}(\textbf{r},t) &= \left( \frac{\hat{\textbf{p}}^{2}}{2m} + \underbrace{v(\textbf{r},t) +  v_{\textrm{Mxc}}([n,q_{\alpha}];\textbf{r},t) }_{v_{\textrm{KS}}([v,n,q_{\alpha}];\textbf{r},t)} \right)\varphi_{i}(\textbf{r},t) , \label{eq:el-pt-td-KS-schroedinger}  \\
\left(\frac{\partial^{2}}{\partial t^{2}} + \omega_{\alpha}^{2}\right) & q_{\alpha}(t) = -\frac{j_{\alpha}(t)}{\omega_{\alpha}} + \omega_{\alpha}\boldsymbol{\lambda}_{\alpha}\cdot \textbf{R}(t) \, , \label{eq:Maxwell-mode-q}
\end{align}
where the total electronic dipole moment can be expressed as $\textbf{R}(t) =  \int d^{3}\textbf{r} \,e\,\textbf{r} \, n(\textbf{r},t)$. The initial conditions of the mode-resolved Maxwell equations (\ref{eq:Maxwell-mode-q}) are $q_{\alpha}(t_{0}) = q_{\alpha}(0)$ and $\dot{q}_{\alpha}(t_{0}) = p_{\alpha}(0)$. We choose the zero-electric field condition for the initial displacement coordinate $q_{\alpha}(t_{0}) = - \left(\boldsymbol{\lambda}_{\alpha}\cdot \textbf{R}\right)/\omega_{\alpha}$. The external potential $v(\textbf{r},t) = v_{\text{ext}}(\textbf{r}) + \hat{\textbf{R}}\cdot\textbf{E}(t)$ has a contribution from the nuclei $v_{\text{ext}}(\textbf{r})$ and $\textbf{E}(t)$ is the dipole-approximated electric field of the laser. The mean-field exchange-correlation potential can be separated into the usual Hartree exchange-correlation potential that is known from electronic TDDFT and an additional photon exchange-correlation potential~\cite{flick2019} as
\begin{align}
v_{\textrm{Mxc}}([n, q_{\alpha}];\textbf{r},t) = v_{\textrm{Hxc}}([n];\textbf{r},t) + v_{\textrm{Pxc}}([n, q_{\alpha}];\textbf{r},t) \, . \label{eq:mean-field-xc-potential-explicitly}
\end{align}
The approximations we apply to the potentials of $v_{\textrm{Mxc}}([n, q_{\alpha}];\textbf{r},t)$ are the adiabatic local density approximation (ALDA) to the Hartree exchange-correlation potential and the photon random-phase approximation to the photon exchange-correlation potential given as follows:
\begin{align}
v_{\textrm{Hxc}}([n];\textbf{r},t) &\simeq \frac{e^{2}}{4 \pi \epsilon_{0}}\int d^{3}\textbf{r}'\frac{n(\textbf{\textbf{r}}',t)}{|\textbf{r} - \textbf{r}'|} + v_{\textrm{xc,ALDA}}([n];\textbf{r},t) \, ,  \nonumber \\
v_{\textrm{P}}([n,q_{\alpha}];\textbf{r},t)&\simeq e^{2}\sum_{\alpha=1}^{M} \left(\int d^{3}\textbf{r}' \boldsymbol{\lambda}_{\alpha} \cdot \textbf{r}'n(\textbf{r}',t) - \omega_{\alpha} q_{\alpha}(t) \right)\boldsymbol{\lambda}_{\alpha} \cdot \textbf{r} \, . \nonumber
\end{align}

To obtain the harmonic spectrum of the coupled system, we need the dipole response. For a laser-driven system, the experimentally measured harmonic spectrum is proportional to the modulus squared of the Fourier transformed laser-induced dipole acceleration~\cite{schafer1997,sundaram1990}:
\begin{align}
H(\omega)	= \left| \int_{0}^{T} dt \frac{d^{2}}{dt^{2}} \textbf{R}(t) \, e^{-i\omega t} \right|^{2} . \label{eq:harmonic-spectrum}
\end{align}
Our description of the coupled system treats the photon field as a dynamical variable of the system such that the Maxwell field couples to the electronic system, which results to a fully self-consistent description of the light-matter system. Similar to Eq.~(\ref{eq:harmonic-spectrum}), we can equally compute the harmonic spectrum due to the photon field. The power spectrum of the displacement field can be determined from the photon displacement coordinate $q_{\alpha}(t)$ using the following expression
\begin{align}
H_{q}^{(\alpha)}(\omega)	= \left| \int_{0}^{T} dt \frac{d^{2}}{dt^{2}} q_{\alpha} (t) \, e^{-i\omega t} \right|^{2} \, . \label{eq:harmonic-spectrum-pt}
\end{align}
This quantity gives direct access to the power spectrum of the displacement field since $q_{\alpha}(t)$ is a mixed matter-photon observable in the length gauge~\cite{rokaj2017}

\section{Harmonic generation with azulene}
\label{app:azulene-harmonic-generation}

In this section, we provide the numerical details of the \emph{ab initio} calculation of the SHG and THG from the azulene molecule. To describe the azulene molecule, we compute the electronic structure using a cylindrical real space grid of $8$ \AA~length with the radius of 6 \AA~in the $x$-$y$ plane and a spacing $\Delta x = \Delta y = \Delta z = 0.16$ \AA. The core electrons of the hydrogen and carbon atoms are described using LDA Troullier-Martins pseudopotentials~\cite{troullier1993} and the $48$ valence electrons that amounts to $24$ doubly occupied Kohn-Sham orbitals are explicitly described in our calculations. 

In our simulations, we induced dynamics in the electron-photon coupled system with a laser pulse whose time-dependent electric field is of the form
\begin{align}
\textbf{E}(t) &= \text{E}_{0}\cos(\omega t)\sin^{2}\left(\pi\frac{t}{\tau}\right) \textbf{e}_{x} \, , \quad 0 \leq t \leq \tau \, . \label{eq:electric-field}
\end{align}
The total pulse duration is fixed to 100 cycles, $T= 100 * 2\pi/\omega$ which corresponds to 170.59~fs, when the fundamental frequency $\omega$ is set to $\omega=\omega_{\pi-\pi^{*}}/2$. The electric field amplitude used for the SHG (THG) is $E_{0}=8.68\times 10^{8}$~V/m ($E_{0}=3.88\times 10^{9}$~V/m) with the laser intensity of $I_{0}=1.0\times 10^{11}$~W/cm$^{2}$ ($I_{0}=2.0\times 10^{12}$~W/cm$^{2}$). The occupied Kohn-Sham orbitals of Eq.~(\ref{eq:el-pt-td-KS-schroedinger}) were propagated self-consistently with the Maxwell equation (\ref{eq:Maxwell-mode-q}) with a time-step of 0.02 a.u. (0.48 attoseconds).

\section{Harmonic generation from photonic degrees}
\label{app:photon-harmonic-generation}

\begin{figure}[bth]
\centerline{\includegraphics[width=0.5\textwidth]{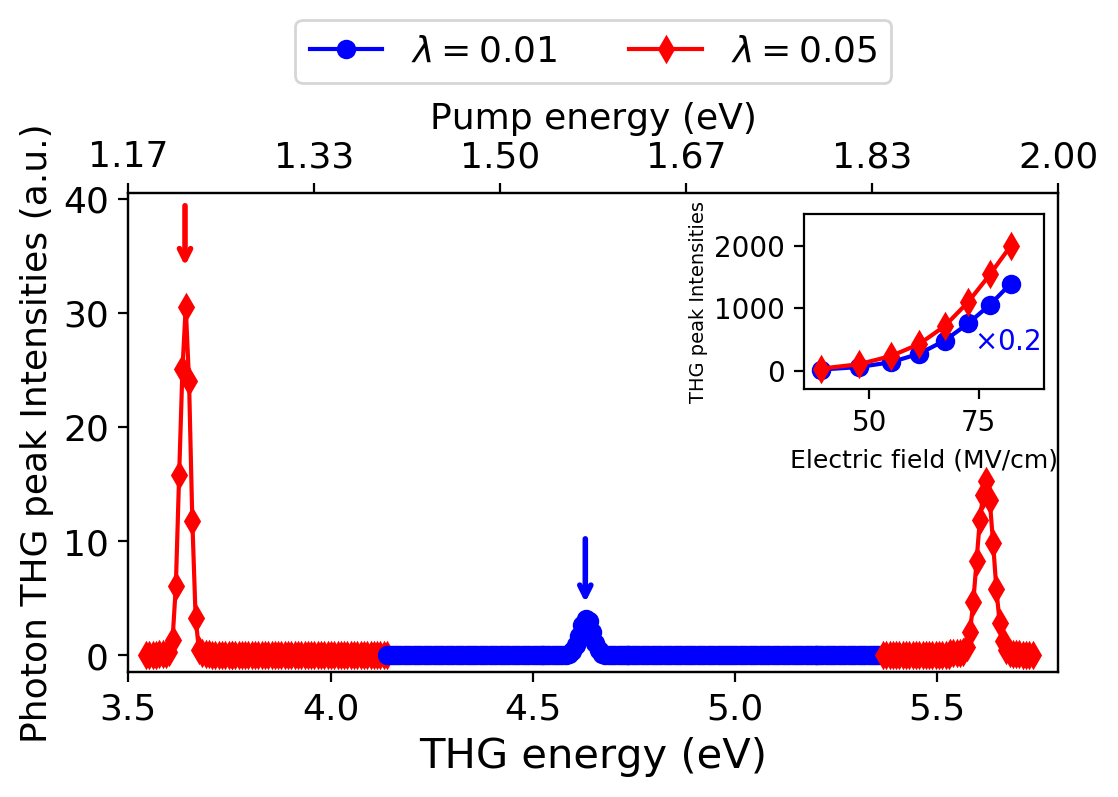}}
\caption{The \textit{photon} THG spectrum computed from the displacement field of the coupled system inside the cavity. For different $\lambda$, THG peaks emerge at the lower and upper polariton states. The maximum intensity of the upper polariton for  $\lambda=0.01$ is 0.0058 a.u. The inset show a cubic dependence of the \textit{photon} THG intensity for different field amplitudes for the peak intensity indicated with arrows.}
\label{fig:photon-THG-omega}
\end{figure}

In this section, we show that the THG process can be computed from the power spectrum formula of the photon displacement coordinate as demonstrated for the SHG case in Sec.~\ref{sec:photon-SHG}. To do this, we perform the same simulations as in Sec.~\ref{subsec:THG-from-polaritons} and calculate the \textit{photon} THG from Eq.~(\ref{eq:harmonic-spectrum-pt}). In Fig.~(\ref{fig:photon-THG-omega}), we show the calculated spectra of the \textit{photon} THG where we find THG peaks from the polariton states as in the case of the dipole radiated field in Sec.~\ref{subsec:THG-from-polaritons}. The inset show the THG intensity as a function of pump field amplitude which has a cubic dependence as expected for THG.

\begin{figure}[bth]
\centering
\begin{minipage}[b]{0.45\linewidth}
	\centerline{\includegraphics[width=8.5cm,height=7.5cm]{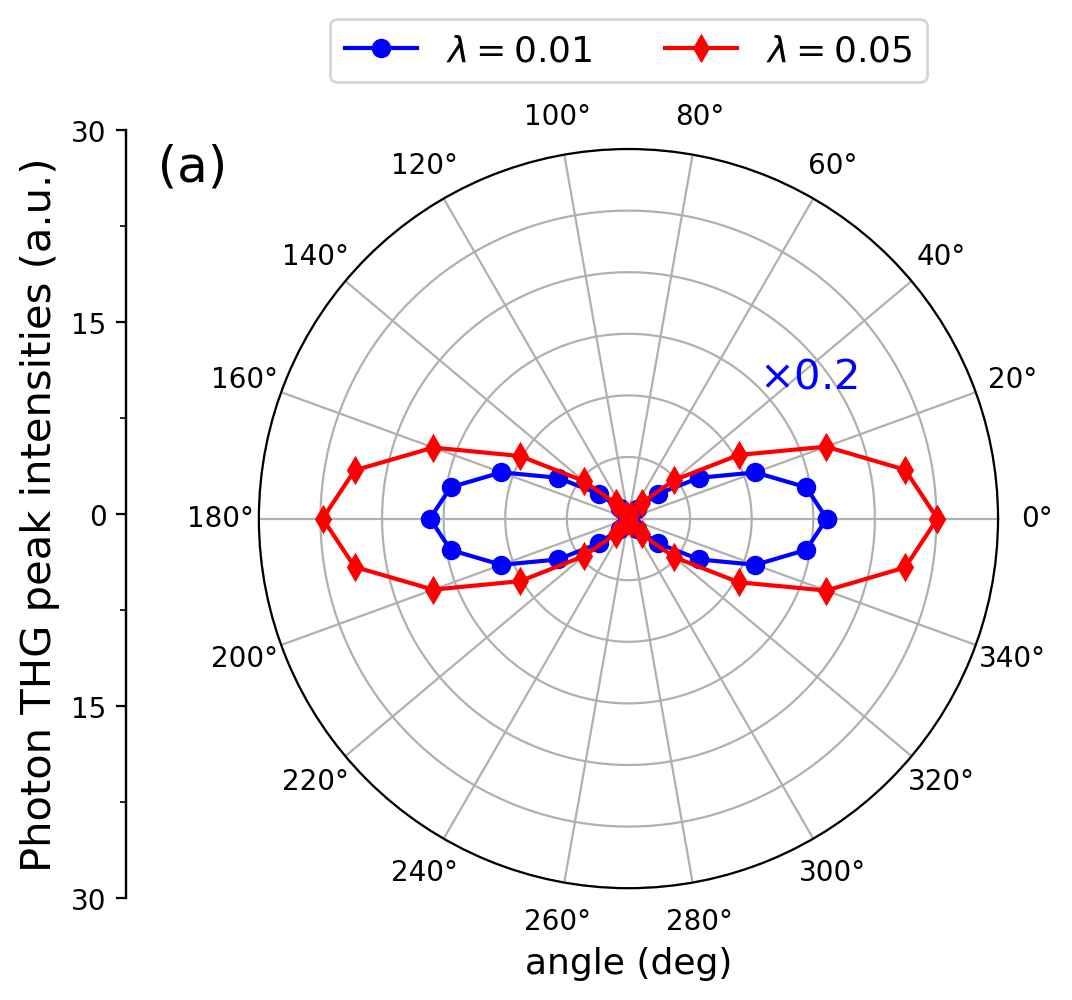}}
\end{minipage}
\\
\begin{minipage}[b]{0.45\linewidth}
	\centerline{\includegraphics[width=8.5cm,height=7.5cm]{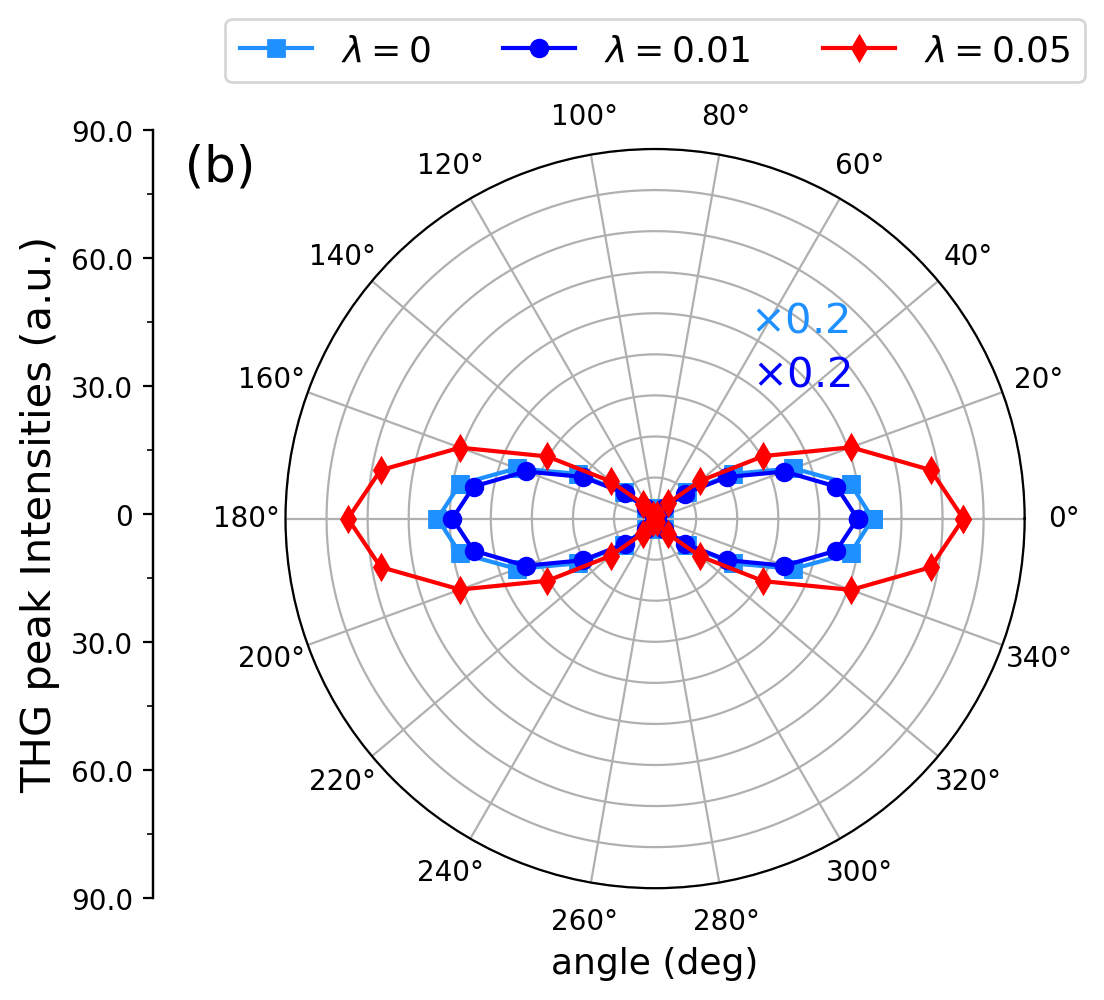}}
\end{minipage}
\caption{(a) Calculated pump-polarization-dependent \textit{photon} third-harmonic radiation pattern obtained from the photon displacement coordinate. (b) THe calculated pump-polarization-dependent third-harmonic radiation pattern obtained from the electronic dipole response.}
\label{fig:pt-THG-polar-spectrum}
\end{figure}

Next, we compute the polarization dependence of the \textit{photon} THG spectra for the central third-harmonic peak of the lower polariton states of Fig.~(\ref{fig:photon-THG-omega}). The polarization angle is chosen as discussed in Sec.~\ref{subsec:SHG-from-polaritons} where we obtain a similar THG emission pattern for the different values of $\lambda$ as shown in Fig.~(\ref{fig:pt-THG-polar-spectrum}).

\vspace{10em}

\bibliography{01_light_matter_coupling} 

\end{document}